# Influence of PEG on the Clustering of Active Janus Colloids

*Mohammed A. Kalil[1], Nicky R. Baumgartner[2], Marola W. Issa[3], Shawn D. Ryan[2,4], Christopher L. Wirth[3,*,†]*

ORCID ID for S. D. Ryan: 0000-0003-2468-1827

ORCID ID for C. L. Wirth: 0000-0003-3380-2029

[1]Department of Chemical and Biomedical Engineering, Washkewicz College of Engineering, Cleveland State University, 2121 Euclid Avenue, Cleveland, Ohio 44115, United States

[2]Department of Mathematics and Statistics, College of Science and Health Professions, Cleveland State University, 2121 Euclid Avenue, Cleveland, Ohio 44115, United States

[3]Department of Chemical and Biomolecular Engineering, Case School of Engineering, Case Western Reserve University, Cleveland, Ohio 44106, United States

[4]Center for Applied Data Analysis and Modeling, Cleveland State University, Cleveland, Ohio 44115, United States

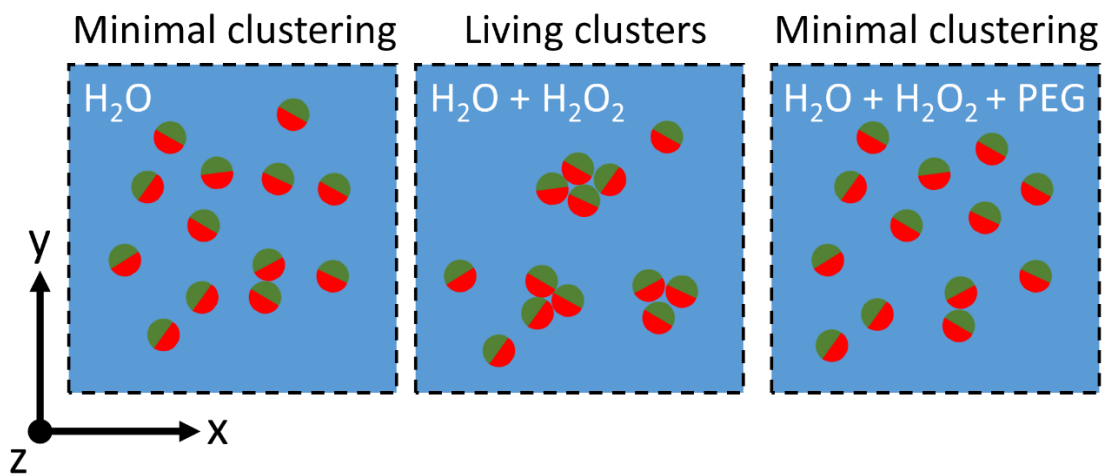

**ABSTRACT**

Micrometer scale "colloidal" particles that propel in a deterministic fashion in response to local environmental cues are useful analogs to self-propelling entities found in nature. Both natural and synthetic active colloidal systems are often near boundaries or are located in crowded environments. Herein, we describe experiments in which we measured the influence of hydrogen peroxide concentration and dispersed polyethylene glycol (PEG) on the clustering behavior of 5 µm catalytic active Janus particles at low concentration. We found the extent to which clustering occurred in ensembles of active Janus particles grew with hydrogen peroxide concentration in the absence of PEG. Once PEG was added, clustering was slightly enhanced at low PEG volume fractions, but was reduced at higher PEG volumes fractions. The region in which clustering was mitigated at higher PEG volume fractions corresponded to the region in which propulsion was previously found to be quenched. Complementary agent based simulations showed that clustering grew with nominal speed. These data support the hypothesis that growth of living crystals is enhanced with increases in propulsion speed, but the addition of PEG will tend to mitigate cluster formation as a consequence of quenched propulsion at these conditions.

1. **INTRODUCTION**

Micrometer scale 'active' Janus particles generate a mechanical force from cues in the environment local to the particle.[1–3] Propulsion generated from the mechanical force in active Janus particles often permits these systems to serve as synthetic analogs to microorganisms that propel by a variety of mechanisms.[4] In both synthetic and biological systems, there is a strong influence of crowding on the individual (*i.e.* near interfaces) and collective (*i.e.* near neighbors) dynamics of active particles.[5–11] Rich phase behavior have been predicted for cases with persistent crowding.[12–14] For instance, a single active Janus particle will propel along a boundary as a consequence of both non-conservative surface interactions and local perturbations in the solute field.[15,16] Further, microorganisms have complex interactions with neighbors or nearby surfaces that are a mix of both conservative and non-conservative interactions.[17–19] Such interactions are central to the initial stages of bacterial biofilm formation.[20] Dynamic collective behavior among crowded swimmers is not restricted to boundaries, self-propelled particles will also tend to cluster with neighbors as a consequence of self-propulsion. Work on self-propelled systems have found that particles tend to form clusters that grow as propulsion speed increases,[21] but there is also a chemotactic influence similar to what was predicted for active particles approaching a boundary.[22] Such phenomena may have significant influence on the formation of microorganism biofilms.[23]

One example of previous work on non-catalytic Janus swimmers suggested clustering is mediated by the nominal speed of the active particles.[21] Graphite capped Janus particles propelled in response to the local demixing of a near critical mixture of water and 2,6-lutidine. Particles were illuminated by a focused beam of 532 nm. The graphite cap absorbed at this wavelength, thereby generating heat that demixed the critical solution and produced a local gradient in solute on the length scale of the particle. The local gradient of solute induced diffusiophoretic motion.

Clustering in ensembles of these particles was studied at varying intensity of illumination, which was used to control the nominal speed of swimming. Particles were observed to cluster to a larger extent at larger nominal speeds. The authors attributed clustering in this instance to a self-trapping mechanism, in which clusters formed from particles colliding with longer time scales for reorientation. Ultimately, particles with larger nominal speed had higher probabilities of collision and thus tended to form larger clusters. This work illustrated that nominal speed of active particles is one key feature for controlling clustering among ensembles of active particles. Speed was controlled via variation in illumination intensity, but other factors may be employed to influence speed to alter the probability of collisions. For instance, our previous work has shown the addition of charged nanoparticles quenches the propulsion speed of active Janus colloids via both an increase in conductivity and the influence of hydrodynamic hindrance in response to depletion attraction [24].

Herein, we describe experiments that show how the addition of polyethylene glycol (PEG) effects clustering of catalytic active Janus particles near a boundary, depending on the volume fraction of added material. First, we found that platinum capped Janus particles in the absence of PEG clustered in response to the addition of hydrogen peroxide. Clusters formed as the concentration of hydrogen peroxide was increased. Second, we found that although clustering was initially slightly enhanced upon addition of PEG, suggesting that PEG acted as a depletant, higher concentrations of PEG similar to those from previous work reduced clustering. Agent-based simulations that complemented these experiments showed that clustering was enhanced (decreased) with increased (reduced) propulsion speed, but still under-predicted the extent of clustering that was observed in experiments. This under-prediction was likely a consequence of the simulation not accounting for both the near-field interactions between particles and also phoretic attraction that arises from perturbations of the local chemical field. Taken together, these results demonstrate the important interplay of conservative and non-conservative interactions on the dynamic behavior of active Janus particles.

## 2. MATERIALS AND METHODS

**2.1. Fabrication of Janus particles and fluid cell assembly.** Platinum-coated polystyrene particles were prepared via glancing angle deposition[25,26]. Briefly, monolayers of 5 μm polystyrene particles obtained from Fisher Scientific were prepared on a 20 mm x 20 mm diced silicon wafer (Ted Pella, INC) with spin coating. High quality monolayers are achieved by carefully selecting the concentration and volume of the suspension deposited onto each wafer. The combination of concentration and volume of the suspension varies with particles size, temperature, and humidity. High coverage of the wafer produces significantly more coated particles per each deposition, thus increasing efficiency. For that reason, a good balance between increasing monolayer coverage and limiting defects is required. Resultant monolayers were analyzed with Scanning Electron Microscopy to ensure high quality coverage.

Next, platinum was deposited on the exposed surface area of particles comprising the monolayer. Physical vapor deposition (PVD) (DV-502A Turbo High Vacuum Evaporator) was used to deposit a cap of 20 nm nominal thickness at a rate of 1 Å/s. We expect particles to have a nominal thickness of 20 nm cap at the 'north pole', but a thickness that decays to 0 nm at the particle equator[27]. After coating, Janus particles were removed from the silicon substrate via bath sonication and stored in ultra-pure water.

Experiments were started by introducing a fixed concentration of Janus particles into a solution of hydrogen peroxide at a defined concentration. For all experiments, Janus particle concentrations were chosen to achieve an area coverage in the region of interest (ROI) of 2% - 3%, corresponding to ~350 particles in a given ROI. This low to intermediate particle concentration has previously been observed to produce clustering in active systems that is dynamic in nature, with clusters reaching an equilibrium size at a given nominal propulsion speed rather than experiencing unbounded growth of gradually larger crystals[30]. Previous work has termed clusters in this regime to be "living crystals".[28] Hydrogen peroxide concentration was changed systematically to test the effects of propulsion speed. Each sample was then added to the fluid cell, which consisted of one microscope slide and one coverslip of size 1 separated by an adhesive spacer of 0.120 mm thickness and 4.5 mm diameter as a viewing window. PEG (Alfa Aesar, Lot #1019473) with a molecular weight MW = 6000 was used as received.

Experiments were conducted by observing particles at 20x magnification for 30 minutes. Note that during an experiment, there occasionally were ROIs in which Janus particle surface concentrations were outside the envelope of 2% - 3% or bubbles had formed. The ROI was moved in these situations during the 30-minute experiment. We did not expect, nor did we observe, evidence of variations in cluster behavior when moving an ROI to an adjacent region because there are no lateral variations in experimental conditions within the fluid cell (beside particle coverage). Thus, results were merged from different ROI's, but in the same fluid cell, with appropriate timing. Note also that bubble formation at hydrogen peroxide concentrations >3% was significant such that propulsion of individual particles and clustering was affected by the presence of bubbles. Thus, the highest peroxide concentration tested herein was 3%.

**2.2. Image processing and analysis.** An upright microscope (Olympus BX51WI) interfaced with a camera (Hamamatsu ORCA-R2 C10600-10B) was used to capture image stacks of particle dynamics. Experiments were typically conducted at frame rates between 8 - 16 frames per second (fps), but were processed in ImageJ to retain one frame every 10 seconds for analysis. Micrographs showed particles as dark with a light background. Further image processing was conducted to measure the size of each object in the ROI such that those objects could be binned into singlets (1 particle), intermediates (2-3 particles), or clusters (4+ particles). This was achieved by first making the image binary such that all particles were black (grayscale value = 0) and a background of white (grayscale value = 255). Although this step in processing is not ideal for tracking small spatial fluctuations, using binary images was effective for counting and binning the size of objects in each frame.

Once a binary image was obtained, the pixel size of each object was measured. An approximate expected pixel size of singlets was first estimated from knowledge of the microscope magnification (0.5119 µm/pixel) and the largest nominal cross-sectional area of a single particle (72 pixels$^2$). However, given that objects appear slightly larger as a consequence of optical effects, we found the real singlet cross-sectional area from an image of known singlets to be slightly larger with a range of 80 pixels$^2$ - 120 pixels$^2$. With knowledge of singlet size, we then binned each object of each retained image into the above noted classifications. The values reported herein for each frame are percent observed, which was calculated as follows:

$$\% \; Observed = \frac{N_i}{\sum_{i=1}^{3} N_i} \tag{1}$$

where $N_i$ is the number of objects observed of that classification in a given frame and $i$ indicates the classification of singlet ($i = 1$), intermediate ($i = 2$), and cluster ($i = 3$). Larger values of percent observed for a given classification implies that a larger fraction of the total number of objects belong to that classification.

**2.3** **Agent-based simulation.** In this section, we briefly review a recent novel agent-based model developed specifically to study active Janus particles[24]. The idea behind the model came from extensive study of active bacteria such as *E. coli* and *B. subtilis* who exhibit similar motion and physical forces. However, Janus particles behave even more like mathematical models due to the enhanced control over their design as opposed to growing bacteria cultures. In this work we use a two-dimensional simplified ODE model relying on overdamped dynamics (e.g., Stokes Law) that accounts for the height above a boundary surface through restricted mobility coefficients with magnitudes determined by the center of mass height. We assume that all the particles are at the same effective height above the bottom surface, $h$, resulting in a quasi-2D modeling approach.

This first principle model is derived by balancing the forces and torques on each Janus particle. The underlying approach is to model a platinum-capped particle as a force dipole with an excluded-volume where the force of self-propulsion due to the chemical reaction balances with the hydrodynamic drag force on the particle body. Here we focus on the net motion which exhibits the same swimming patterns as bacteria [29–31]. Since we study a large number of simple particles and their resulting motion, the exact form of an individual particle is relaxed for computational efficiency.

The key assumption in the model is that we are in a low Reynolds number regime, consistent with the Janus particle size, swim speed, and fluid viscosity. Each of the $N$ identical particles obeys the following system of equations

$$\dot{\mathbf{x}}_i = q_{xy}\left(v_0 \mathbf{d}_i + \mathbf{u}(\mathbf{x}_i,t) + \frac{1}{6\pi\eta a}\sum_{i\neq j}^{N} \mathbf{F}(\mathbf{x}_i - \mathbf{x}_j) + \sqrt{D_t}\frac{dW}{dt}\right) \quad (2)$$

$$\dot{\mathbf{d}}_i = q_\theta\left(-\mathbf{d}_i \times [\nabla \times \mathbf{u}] + \sqrt{D_\theta}\frac{dW}{dt}\right) \quad (3)$$

Using this system one can track the center of mass and the orientation of each particle, which are crucial in understanding the dynamics of these interactions. Furthermore, observe that all the interactions between particles are pairwise. This is due to the linearity of the Stokes flow in the low Reynolds number regime and the fact that interparticle interactions are accounted for using a pairwise sum is consistent with the *semi-dilute* regime of a moderate volume fraction suspension of particles. The first equation governs the translational velocity of each particle with translational hindrance coefficient $q_{xy}$. The hindrance simulates a thin-film three-dimensional domain using a two-dimensional modeling approach. The first term allows for self-propulsion of magnitude $v_0$ in the direction of its orientation $\mathbf{d}_i$ driven by the chemical reaction with the hydrogen peroxide that induces the propulsion. The second term is the local fluid velocity at the location of the $i^{th}$ particle capturing the hydrodynamic interactions with a particle's surroundings as a function of position $\mathbf{x} \in R^2$ at time $t$ generated by a Janus particle at the origin with orientation $\mathbf{d}_i$ is given by

$$\mathbf{u}(\mathbf{x},t) = -\frac{U_0(a+h)}{3\pi} \sum_{i=1}^{N} \{\nabla^3[\log(|\mathbf{x} - \mathbf{x}_i(t)|)] \cdot \mathbf{d}_i(t)\} \mathbf{d}_i(t) \qquad (4)$$

where $U_0 = \zeta \eta \ell^2 v_0 > 0$. This represents the dipole moment following Stokes drag law, where $\zeta$ is a shape coefficient, $\eta$ is the ambient fluid viscosity, $v_0$ is the isolated swimming speed of the Janus particle of diameter $\ell$ [30]. Since $a$ is the particle radius, then $a + h$ is the height of the center of mass above the bottom surface. The third term is an interaction potential that enforces the excluded-volume of the particle body and provides an intermediate range attraction exhibited in clustering regimes. The last term is a translational diffusion modeled as white noise represented as the derivative of a Weiner process with strength $D_t$.

The primary change is the excluded-volume interaction modeled by the force $\mathbf{F} = -\nabla U(\mathbf{x})$ which is the gradient of a Morse potential that has successfully been used to capture typical clustering behavior in self-propelled particles in recent works [32]. The generalized Morse potential is given by

$$U(\mathbf{x}_i) = \sum_{j \neq i} [C_r e^{-\frac{|x_i - x_j|}{\ell_r}} - C_a e^{-\frac{|x_i - x_j|}{\ell_a}}] \qquad (5)$$

Here $\ell_a$ and $\ell_r$ represent the effective distances of the attractive and repulsive potentials respectively. The coefficients $C_a$ and $C_r$ govern the relative amplitudes of each piece of the potential. Thus, each particle has some interaction attractive/repulsive with all the other particles determined by their interparticle distance. The long-range hydrodynamic interactions governed by $\mathbf{u}(\mathbf{x}, t)$ initially attract neighboring particles, but the Morse potential reinforces the clustering observed in experiment. In fact, an early version of the model with a purely repulsive potential, such as the Yukawa potential used in [24], allow for the initial formation of a cluster, but the components soon break apart without the additional reinforcing interaction. As previously demonstrated, the Yukawa potential is suitable for dilute regimes where motion is dominated by self-propulsion and the Morse potential is better suited for semi-dilute regimes where interparticle interactions dominate motion at moderate concentrations.

The second equation accounts for the balance of torques on each Janus particle with a corresponding angular hindrance $q_\theta$. The first term is analogous to the classical Jeffery's equation that allows the orientational dynamics of the point dipole Janus particles to interact with the surrounding fluid as prolate ellipsoids [33]. Since the model is two-dimensional the rotational vorticity portion of Jeffery's equation is zero (acts entirely out of the plane of the computational domain) and all that remains is the term due to the rate-of-strain. The last term represents the rotational diffusion of the Janus particle with strength $D_\theta$. In the angular component the hindrance $q_\theta$ reduces the angular velocity of the particle as it approaches the bottom surface. **Table 1** summarizes simulation conditions used herein.

| Simulation Conditions | |
|---|---|
| particle radius (μm) | 2.5 |
| temperature (K) | 298 |
| fluid viscosity (Pa s) | 0.00089 |
| particle density (kg/m$^3$) | 1055 |
| translational diffusion coefficient in bulk (m$^2$/s) | 9.80998 x 10$^{-14}$ |
| rotational diffusion coefficient in bulk (r$^2$/s) | 0.01177198 |
| total number of frames (#) | 1800 |
| time step (s) | 0.1 |
| dipole moment, $U_0$ | 0.01 |
| Morse potential parameter $C_r$ | 0.60 |
| Morse potential parameter $C_a$ | 1.00 |
| Morse potential parameter $l_r$ | 0.50 |
| Morese potential parameter $l_a$ | 1.00 |

**Table 1: Simulation conditions.**

## 3. RESULTS AND DISCUSSION

**3.1. Cluster formation depended on peroxide concentration.** The ensemble dynamics of platinum capped Janus particles were initially measured in a 3% hydrogen peroxide solution. We observed dynamic clustering, in which individual particles joined and departed a variety of intermediate and cluster structures during the experiment. For example, a single particle would typically pair with one, two, three, or more particles, but then escape those particles for finite periods of time during an experiment, only to reunite with different partners later in the experiment (see **Fig. 1**). Such behavior is characteristic of living crystals, in which an active system will form crystals that experience dynamic exchange, rather than gradually form one large cluster as one would expect with an attractive system. Previous work has provided strong evidence for the mechanism by which dynamic clustering arises in active systems [34,35]. Living crystals arise from the dynamic exchange of active particles between clusters of finite size and the bulk. Particles collide at a given frequency because of directional activity, but then reorient as a consequence of random orientation fluctuations. Reorientation of the directionality of the active particle causes it to either leave or restructure the cluster. The result of such behavior is that the mean dynamic cluster size will increase with increased velocity. Structures formed in our experiments were typically small, normally not exceeding ~6 particles. Such small size is consistent with previous work showing the mean size of living crystals grows with propulsion speed of individual active particles.

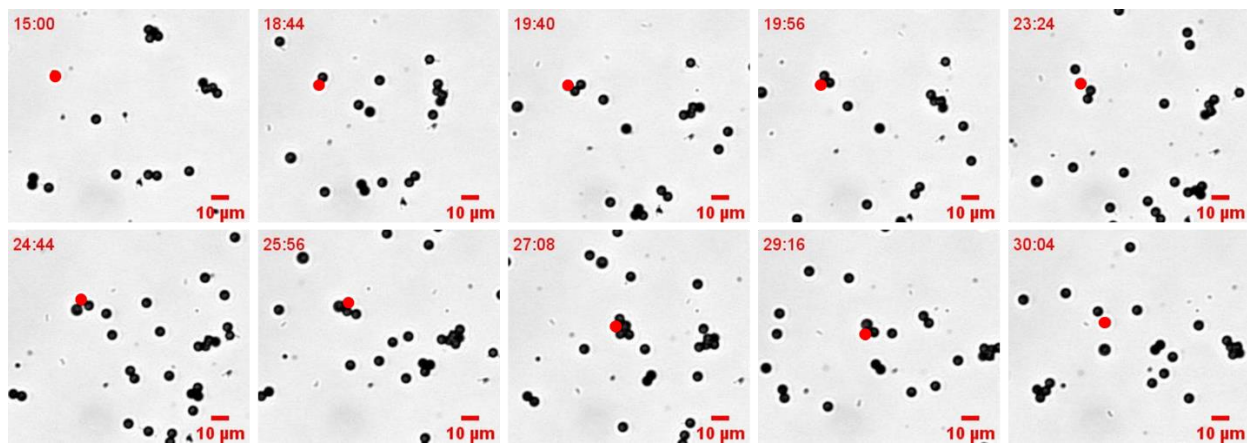

**Figure 1: Tracking the structures sampled by a single particle.**
A single particle (covered by a red dot) was tracked for ~15 minutes during an experiment. The particle sampled many different structures, including intermediate (2 – 3 particles) and cluster (4+ particles) structures. This dynamic behavior was characteristic of all our experiments.

We measured the evolution of singlets (1 particle), intermediates (2 and 3 particle dimers/trimers), and clusters (4+ particles) as a function of time. The fraction of clusters increased in time (see **Fig. 2(c)**) at the expense of intermediates and singlets (see **Figs. 2(a)** and **2(b)**) for Janus particles in a 3% solution of hydrogen peroxide. Data shown in **Figure 2** is from a single realization of our experiments. There were significant temporal fluctuations in measured values as a consequence of the dynamic process unfolding during a single experiment. The process proceeded with singlets merging and separating, which contributed to the temporal fluctuations in both of these measured quantities. Clusters ($\geq$ 4 particles) were more robust against temporal fluctuations because individual merging or separation events do not necessarily alter the 'cluster' classification.

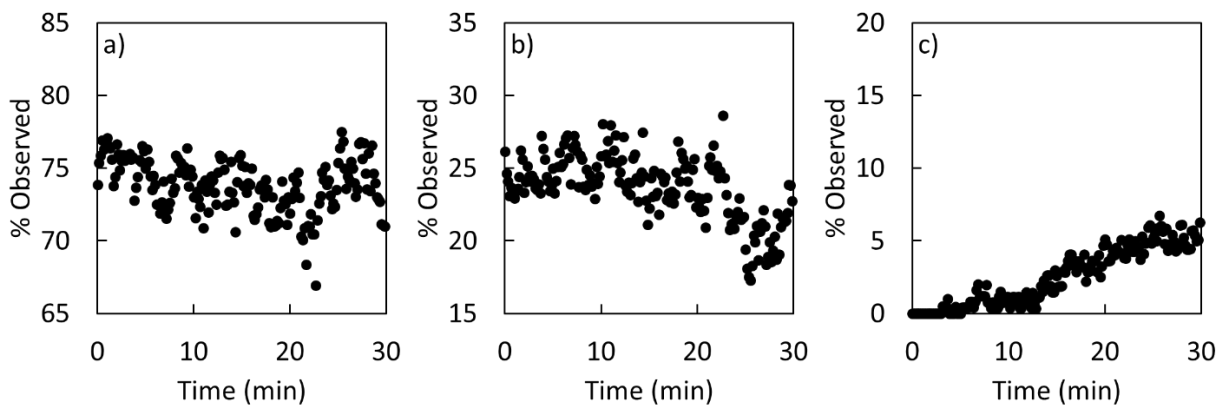

**Figure 2: Evolution of (a) singlets (1 particle), (b) intermediates (2 and 3 particle dimers and trimers), and (c) clusters (4+) for active Janus particles in 3% hydrogen peroxide.** A slight decrease in singlet percentage was observed, along with a stronger decrease in intermediates upon formation of clusters. Measurements of singlets and intermediates had strong temporal fluctuations because of the dynamic merging and separating process.

Singlet percentage did not decrease more strongly in response to the formation of intermediates and clusters for a single experimental realization (see **Figs. 2(a)**). The primary reason for the singlet observed percentage remaining steady for the beginning of each experiment was that singlets entered from above the region of interest. The number of singlets entering the frame in the first ~10 min of a single experimental realization was significant and these additional particles keep the percentage of singlets from dropping at an early time when viewing the entire time window of the experiment. Following the initial changes in nominal singlet concentration, clustering reached a dynamic equilibrium after ~10 minutes. We accounted for these phenomena by reporting dynamic equilibrium measurements obtained by taking the mean over ~5 minutes in the plateau region indicative of this state (see *Supporting Information*). Further, averaging over many realizations of experiments with the same conditions shows a significant trend in singlets drop once the system reaches dynamic equilibrium. Such behavior is consistent with dynamic cluster formation where single active Janus particles initially form dimers, then trimers, then finally clusters of at least four particles.

Clustering of active Janus particles was tracked for systematic variations in hydrogen peroxide concentration between 0% - 3%. Concentrations of hydrogen peroxide over this range for the same system have previously been found to drive active Janus particles to swim at speeds of up to approximately 1 μm/s.[24] Although such speeds are slower than what has been reported for other systems at similar concentrations, the particles used herein are larger than those systems. Given that swim speed has been shown to decrease with particle radius,[36] speeds for the current system met expectations. The formation of clusters (4+ particles) increased with hydrogen peroxide concentration. Cluster formation over 30 minutes was small in solutions <1.0%, but occurred more strongly at hydrogen peroxide concentrations >1.0%.

Individual realizations were repeated to obtain the average ensemble response as a function of hydrogen peroxide concentration. **Figure 3** shows singlet, intermediate, and cluster observed percentage measured at dynamic equilibrium as a function of the apparent propulsion speed measured in an earlier contribution.[24] There was strong singlet loss coupled with a strong increase in intermediate formation. Minimal clustering was formed at low hydrogen peroxide concentrations (*i.e.* small speeds) while clustering was enhanced at larger hydrogen peroxide concentrations (*i.e.* large speeds). As noted, previous work on active systems showed that cluster size increased with increased propulsion speed at intermediate concentration.[21]

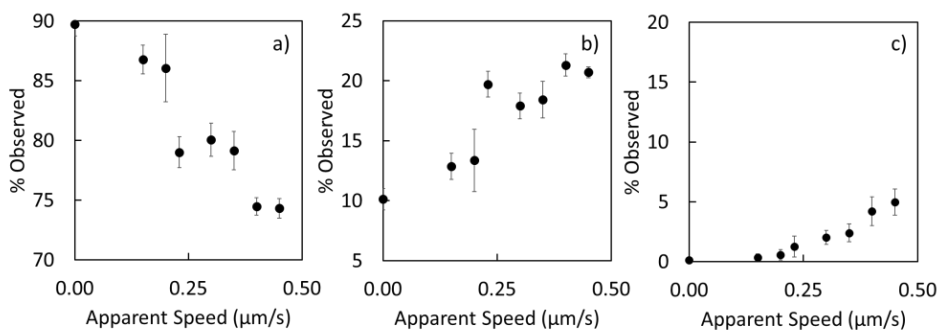

**Figure 3: Mean extent of (a) singlets, (b) intermediates, and (c) clusters at dynamic equilibrium as a function of apparent speed of individual particles.** Similar to Figure 2, these data show the rate of clustering was enhanced as hydrogen peroxide concentration (*i.e.* apparent speed) increased. Error bars are the standard error associated with a given set of experiments.

Active particles propel as a result of a hydrogen peroxide decomposition reaction on the platinum cap that produces a gradient in solute on the length scale of the particle. As has been noted in the literature,[16,37–42] the solute concentration gradient itself in the local environment contributes to clustering in a way that supplements the mechanism of collision and reorientation. This phoretic mechanism arises from the gradient in solute being spatially altered by the nearby boundary. The combination of those two mechanisms – one hydrodynamic and one phoretic in nature – leads to a variety of states experienced by an active particle near a boundary[41]. Note also that near field conservative interactions, including electrostatics and van der Waals attraction may contribute to the longevity of clusters. Moreover, the chemical reaction on the particle surface is accompanied by significant local changes in pH, which would in turn contribute to changes in the surface charge of individual particles. However, given our observations that Janus particles, even in the presence of peroxide, remain mobile near a boundary suggests these changes may not be significant to the extent that surface charge appreciably changes.

**3.2. Impact of PEG on clustering.** Our previous work has shown the addition of either charged nanoparticles or polymer reduced the propulsion speed of active Janus particles.[24] Herein, polyethylene glycol (PEG) of low molecular weight (6K) was added at systematically varied volume fractions equal to and below that of our previous work. Our hypothesis motivating these experiments was the reduction of propulsion speed at fixed hydrogen peroxide concentration of active Janus particles is a consequence of PEG acting as a depletant. The reduction of propulsion speed will reduce the extent to which an ensemble clusters.

**Figure 4** shows cluster observed percentage as a function of PEG concentration for systems at fixed peroxide concentrations of 3% and 0%. Clustering initially was slightly enhanced at small PEG concentrations in the presence of fuel, but was diminished as PEG concentration grew. Enhancement of clustering at small PEG concentration was likely a consequence of increased attractive particle-particle interactions that led to longer lived clusters. This attractive interaction would arise in the event PEG acted as either a depletant or as an avenue for bridging flocculation.

At small PEG concentrations, the apparent speed of Janus particles would be roughly unchanged, thereby producing no change in the collision probability. However, once a collision occurred, the probability of remaining clustered was increased because of the conservative attraction between particles introduced by the presence of PEG, in agreement with PEG acting as a depletant.[43] As PEG concentration was increased to a regime in which Janus particle speed was observed to decrease, the extent of clustering was diminished. Janus particles in the absence of fuel experienced low levels of clustering that were slightly enhanced from that of Janus particles in the absence of both fuel and PEG.

Note the shaded grey area in **Figure 4** corresponds to the PEG regime where propulsion speed was observed to be reduced.[24] A decrease in nominal Janus particle propulsion speed reduced the probability of collision, thereby reducing the extent to which clusters were able to form. Note, the higher PEG concentration would also increase the conservative attractive interaction between particles. However, such attraction is on a sufficiently small length scale that the interaction becomes irrelevant for small collision probabilities. We approximated the length scale of the conservative attraction as being equal to the cube root of the excluded volume of a particle, which included the volume of the shell surrounding the particle with a thickness equal to the radius of gyration of the 6K PEG. For our system, this distance is ~4 μm. At the highest PEG concentration, the probability of collision is exceedingly small because active Janus particles have had speed significantly quenched. Thus, active Janus particles at this concentration are unlikely to enter the attraction region of a neighboring particle.

A secondary effect that reduced the extent to which clusters formed was encountered at the highest PEG concentration. In the worst case scenario observed only at the highest PEG volume fraction, as many as 30% of particles were immobilized on the substrate by the end of the experiment, thereby drastically reducing the probability of collisions. Note that deposition of active Janus particles was only observed at the highest PEG concentration; collision probability was altered only because of the addition of PEG at all other concentrations. Finally, one additional effect could be that of viscosity. The addition of PEG to solution at these conditions may increase the viscosity of the solution by as much as 10%,[44] at the highest PEG concentration, with the lowest three concentrations of PEG having viscosity increases below 1%. Although such change in viscosity is appreciable, it would be insufficient to be wholly responsible for the diminished clustering observed herein.

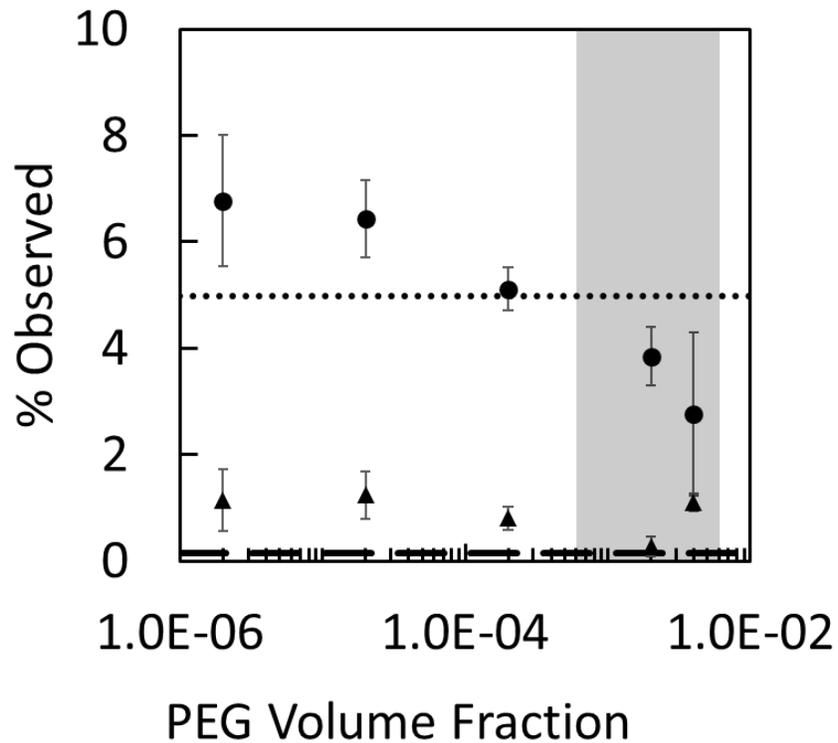

**Figure 4. Extent of clustering at dynamic equilibrium for Janus particles in 3% (circles) and 0% (triangles) hydrogen peroxide with systematic variation in PEG concentration.** The grey region signifies the PEG concentration in which Janus particle propulsion speed was observed to be quenched by the presence of PEG. The dotted line is the % observed clusters in the absence of PEG with 3% hydrogen peroxide, while the broken line is the % observed clusters in the absence of both PEG and hydrogen peroxide. Janus particles had roughly the same propulsion speed (*i.e.* roughly the same collision probability) at lower PEG concentrations, but an enhanced attractive interaction that increased the probability of long lived clusters. At larger PEG concentrations, the propulsion speed

decreased, thereby decreasing the probability of collisions. The increased particle-particle attractive interaction was irrelevant at conditions in which the Janus particles experienced a smaller collision probability. Further, there was very minimal clustering among Janus particles in the absence of hydrogen peroxide as compared to Janus particles in the presence of hydrogen peroxide. Error bars are the standard error associated with a given set of experiments.

Experimental results summarized above reveal a picture consistent with previous work from our own group and others. Janus particles tend to form clusters as propulsion speed increases. Upon the addition of PEG, clustering was initially enhanced at low concentrations, but diminished at higher PEG concentrations where the Janus particle propulsion speed is expected to be low. Taken together, these results reveal that tuning of clustering can be achieved by controlling the collision probability via the propulsion speed. The essence of our findings is that the addition of a PEG achieves a similar effect with respect to clustering as a reduction in peroxide concentration.

Although these data and our previous work forms a strong connection between the addition of co-solute (PEG or charged nanoparticles), the molecular mechanism by which propulsion speed is reduced is unknown. There is strong evidence the mechanism is that of depletion induced attraction between either neighboring particles or the particle and nearby surface. However, it is also possible that adsorption of PEG to the exposed PS face of the Janus particle alters the reaction mechanism in some fashion. Nevertheless, we further tested the relationship between propulsion speed and clustering via agent-based simulations.

## 3.3. Agent-based simulations of clustering as function of nominal speed.

The simulations were run at two area fraction regimes using $N = 350$ (approximately 1.75% area fraction) and at $N = 175$ particles (approximately .89% area fraction) in a semi-dilute regime. The semi-dilute regime is characterized by area fractions where pairwise hydrodynamic interactions are accounted for in particle motion. This is in contrast to the dilute regime where the motion of all particles can be treated as a single particle in an infinite domain. More important than the number of particles is the area fraction, $\rho = N\pi\ell^2/4L^2$, which is kept at or below 5%. The computational domain is taken to be $L \times W$ where $L = 140\ell$ and $W = 110\ell$ where each dimension is roughly 100 times the size of an individual Janus particle with periodic boundary conditions in the $x$ and $y$ directions. This greatly reduces any finite size effects that may be present in other computational approaches.

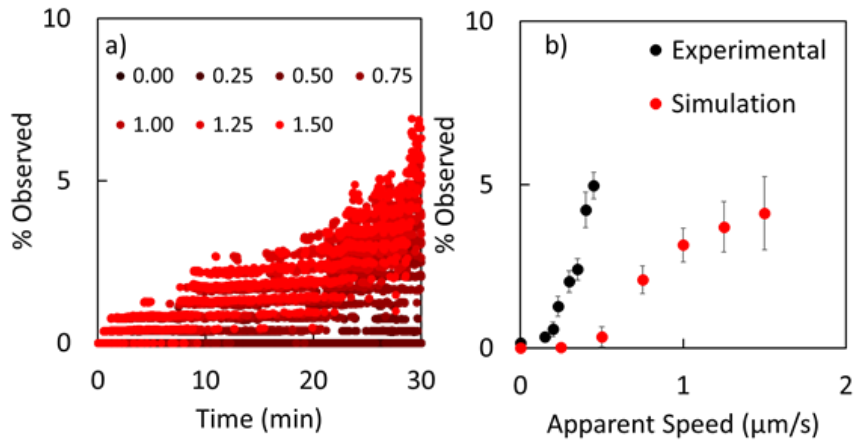

**Figure 5: (a) Simulated cluster observed percent as a function of time for systematic variation in swim speed (units of μm/s) and (b) extent of clustering after 30 minutes in both simulation and experiment for Janus particles as a function of apparent propulsion speed.** The intensity of red in (a) is proportional to the propulsion speed. As observed in experiments, the extent of

clustering increased as a function of both time and propulsion speed.

(b) Simulated clustering under-predicted experimental results even though the simulations were conducted at larger area fractions. One potential explanation for this discrepancy was in the tuning of interparticle potentials on two fronts. First, simulations utilized an approximate expression for interparticle interactions, which could potentially have under-predicted the magnitude attractive conservative interactions, such as van der Waals. Second, the simulation did not take into account chemotactic contributions to clustering. Note the error bars for experimental data in (b) is the standard error for a set of experiments, whereas the error bars for simulated data in (b) is the standard deviation of cluster size for one simulation.

Clustering phenomena generated from the agent-based simulation qualitatively matched that of experimental results (see **Fig. 5**) with the percent of observed clusters increasing as a function of time. Dynamic exchange (i.e. living crystals) occurring between singlets and intermediates, as observed in experiments, was also observed in these simulated data. Further, the extent of clustering increased as a function of apparent propulsion speed. Larger propulsion speeds tended to induce more cluster formation, after some critical speed. Critical phenomena associated with cluster production has previously been observed with active systems.[21] However, there was a quantitative discrepancy between simulations and experiments.

Simulations under-predicted cluster formation as compared to experimental observations (see **Fig. 5(b)**). Previous work has also suggested that interparticle interactions could be generated by chemical gradients,[22] which were not accounted for herein. Phoretic attraction of the same magnitude and supplemental to attraction solely from hydrodynamics could potentially account for the differences between experiments and simulation highlighted in **Figure 5**. Further, the particle-particle interaction utilized for the simulations was an approximation. Although appropriate for gaining insight into trends associated with the phenomena, such an approximation lacks details associated with van der Waals attraction. Increasing van der Waals attraction would lead to longer lived clusters that would tend to enhance the observed cluster percentage in simulations. Finally, in the simulation all particles are assumed to be at the same height $h$ whereas in simulations there is some variability. The predictions of the model increase in accuracy as the number of particles increase and the effects of individual microscopic interactions are minimized.

## 4. CONCLUSIONS

We summarized experiments and simulations in which we measured the influence of variable propulsion speed on the extent and dynamics of clustering of active Janus particles. Ensemble clustering of 5 μm catalytic active Janus particles was measured in response to variations in hydrogen peroxide concentration and dispersed polyethylene glycol (PEG). We found the extent to which clustering occurred grew with hydrogen peroxide concentration in the absence of PEG, was further enhanced at small concentrations of PEG, and was reduced at higher PEG volumes fractions. The region in which clustering was mitigated at higher PEG volume fractions corresponded to the region in which propulsion was previously found to be quenched. We conclude first from these data that increases in collision probability for catalytic active Janus particles will increase cluster formation. Second, these data suggest PEG enhanced attraction between particles to increase cluster longevity at low concentrations, but then reduced cluster probability via a reduction in collision probability. Further, comparing our experimental results to agent based simulations that only consider hydrodynamics and model inter-particle interactions highlights the importance of including other critical features that play a role in active particle clustering.

**AUTHOR INFORMATION**

Corresponding Author

*E-mail: wirth@case.edu

Present Addresses

†Department of Chemical and Biomolecular Engineering, Case Western Reserve University, Cleveland, Ohio 44106

Author Contributions

The manuscript was written through contributions of all authors. All authors have given approval to the final version of the manuscript.

**ACKNOWLEDGMENT**

This work was supported by the Cleveland State University Office of Research Startup Fund, a Faculty Research Development Grant, and the National Science Foundation CAREER Award NSF no. 1752051. The SEM used herein was obtained with support by the National Science Foundation under Grant No. 1126126.